\documentstyle[12pt]{article}
\begin{document}
\title{An electroweak model without Higgs particle}
\author{{Ning Wu}
\thanks{email address: wuning@tofj1.ihep.ac.cn}
\\
{\small CCAST (World Lab), P.O.Box 8730, Beijing 100080, P.R.China}\\
{\small and}\\ 
{\small Division 1, Institute of High Energy Physics, P.O.Box 918-1, 
Beijing 100039, P.R.China}
\thanks{mailing address}}
\maketitle
\vskip 0.8in

\vskip 0.8in

\noindent
[Abstract] In this paper, applying general gauge field theory, we will 
construct an electroweak model. In this new electroweak model, Higgs 
mechanism is not used,  so no Higgs particle exists in the  model. In order 
to keep the masses of intermediate gauge bosons non-zero, we will introduce 
two sets of  gauge fields. We need a vacuum potential to introduce 
symmetry 
breaking and to introduce  the masses of all fields. Except for those terms 
concern of Higgs particle, the fundamental dynamical properties of this 
model 
are similar to those of the standard  model. And in a proper limit, this model 
will approximately return to the standard model. 
\\

\newpage

\section{Introduction}
~~~~ After Yang and Mills founded non-Abel gauge field theory in 1954 
\lbrack 1 \rbrack, gauge field theory has been extensively applied to 
elementary particle theory. Now,  it is clearly known that four kinds of 
fundamental interactions, i.e. strong interactions, electromagnetic 
interactions, 
weak interactions and gravitation, are all gauge interactions, and they can be 
described by gauge theory.  From theoretical point of view, the requirement 
of  gauge invariant determines the forms of interactions. In other words, the 
principle of local gauge invariant plays a fundamental role in particles' 
interaction theory.  \\

~~~~ But for Yang-Mills gauge theory, if a system has strict local gauge 
symmetry, the masses of  gauge fields must be zero. But physicists found 
that 
the masses of intermediate gauge bosons are very large in the forties \lbrack 2 
\rbrack. After introducing the concept of spontaneously symmetry 
breaking and Higgs mechanism which make the gauge fields obtain masses, 
Glashow \lbrack 3 \rbrack, 
Weinberg \lbrack 4 \rbrack and Salam \lbrack 5 \rbrack  founded the well-
known unified electroweak standard model. The standard model is consonant 
well with experiment and the intermediate gauge bosons predicted by the 
standard model have already been found by experiment \lbrack 6 \rbrack, but 
the Higgs particle necessitated by the standard model has not been found by 
experiment until now. Whether Higgs particle exists in nature? If there were 
no Higgs particle, how should we construct the unified electroweak model?  
\\

~~~~ The general gauge field theory  was put forward by Wu recently  
\lbrack 7-8 \rbrack. The main difference between the general gauge field 
theory and 
Yang-Mills gauge field theory is that, there exist massive force-transmitting 
vector fields in the general gauge theory under the precondition that the 
system has strict local gauge symmetry. This characteristic of the general  
gauge theory makes it possible for us to directly construct electroweak model 
without using Higgs mechanism  \lbrack 9 \rbrack. Because Higgs 
mechanism is no longer needed in the new electroweak model, there  exists 
no Higgs particle in the new theory. Furthermore, in a proper limit, the new 
electroweak  model will approximately return to the standard model . So, we 
could 
anticipate that there will exist no contradictions between the  new 
electroweak model and experiments which we will discuss in details later. 
\\

~~~~ In this paper, using the general gauge field theory, we will construct a 
new unified electroweak model. First, we will give the lagrangian for 
electroweak interactions of leptons. In order to introduce symmetry breaking 
of the model and the masses of all fields, we  need a scalar potential which 
we will name it vacuum potential. The functions of the vacuum potential are 
similar to those of Higgs scalar field, but vacuum potential has essential 
differences from Higgs scalar field. After symmetry breaking,  fermions and 
gauge bosons obtain masses. The electroweak interactions of quarks will also 
be discussed in this paper. The dynamical characteristics of this new 
electroweak model are similar to those of the standard model. And in a proper 
limit, except for those  terms concern of Higgs particle, the new electroweak 
model will approximately return to the standard model. In this new 
electroweak model,  we 
will introduce two sets of gauge fields. After some field transformations, one 
set of 
gauge fields will obtain masses and another set will keep massless. The 
existence of  these massless gauge bosons is an important characteristic of 
the new model. If electroweak model should be a minimum model, 
those massless gauge bosons and Higgs particle can not exist in the same 
theory. But if we do not add  the restriction of minimum model to 
electroweak interactions, those massless gauge bosons and Higgs particle can 
exist in the same theory. Therefore, the existence of  these massless 
gauge bosons is not enough to say 
that Higgs particle does not exist in nature for certain. The existence of these 
massless gauge bosons only means that Higgs particle may not exist in 
nature. 
An important prediction of this new electroweak model is that there exists a 
new long-range force in nature which may have applications in the future. At 
the end of this paper, we will present some discussions on some fundamental 
problems of the new electroweak model. 
\\

\section{The lagrangian of the model (leptons) }

~~~~ Up to now, physicists have found that there exist three generations of 
leptons and quarks in the nature. In this chapter, we will discuss the 
electroweak interactions of leptons. For the sake of convenience, let's $e$ 
represent leptons $e,\mu$ or $\tau$, and $\nu$ represent the corresponding 
neutrinos $\nu _e, \nu_{\mu}$ or $\nu_{\tau}$. According to the standard 
model,  $e$ and $\nu$ form left-hand doublet $\psi_L$ which has 
$SU(2)_L$ symmetry and right-hand singlet $e_R$. Neutrinos have no 
right-hand singlets. The definitions of these states are:
$$
\psi_L =\left ( 
\begin{array}{c}
\nu  \\
e
\end{array}
\right )_L
~~~,~~~ y= -1
\eqno{(2.1)} 
$$
$$
e_R
~~~~~~,~~~ y= -2 ,
\eqno{(2.2)} 
$$
where $y$ represents the quantum number of weak hypercharge $Y$.  \\

~~~~ The symmetry of the theory is supposed to be the group $SU(2)_L 
\times U(1)_Y$. The generators of $SU(2)_L$ are denoted as $T_i^L = \tau 
_i/2$ ($\tau_i$ are Pauli matrices), and the generator of $U(1)_Y$ group is 
denoted as $Y$. The electric charge of a particle is determined by Gell-
Mann-Nishijima rule:
$$
Q = T_3^L + \frac{Y}{2}.
\eqno{(2.3)} 
$$

~~~~ Four gauge fields are needed in the new electroweak theory. They are 
two non-Abel gauge 
fields $F_{1 \mu}$ 
and $F_{2 \mu}$ corresponding to the $SU(2)_L$ symmetry and two Abel 
gauge fields $B_{1 
\mu}$ and $B_{2 \mu}$ corresponding to the $U(1)_Y$ symmetry. Two 
$SU(2)_L$ gauge fields 
$F_{1 \mu}$ and $F_{2 \mu}$ can be expanded as:
$$
F_{m \mu} = F^i_{m \mu} \frac{\tau_i}{2}
~~,~~~~(m=1,2)
\eqno{(2.4)} 
$$
where $F_{m \mu}^i~(i=1,2,3)$ are component fields. Corresponding to 
four 
gauge fields, we will 
introduce four gauge covariant derivatives:
$$
D_{1 \mu} = \partial_{\mu} - i g F_{1 \mu}
\eqno{(2.5)} 
$$
$$
D_{2 \mu} = \partial_{\mu} + i g {\rm tg} \alpha F_{2 \mu}
\eqno{(2.6)} 
$$
$$
D_{3 \mu} = \partial_{\mu} - i g' B_{1 \mu} \frac{Y}{2}
\eqno{(2.7)} 
$$
$$
D_{4 \mu} = \partial_{\mu} + i g' {\rm tg} \alpha B_{2 \mu} \frac{Y}{2},
\eqno{(2.8)} 
$$
where $\alpha$ is a dimensionless constant. The strengths of four gauge 
fields are respectively 
defined as:
$$
F_{1 \mu \nu} = \partial _{\mu} F_{1 \nu} - \partial _{\nu} F_{1 \mu} 
- i g \lbrack F_{1 \mu} ~~,~~    F_{1 \nu} \rbrack ,
\eqno{(2.9)} 
$$
$$
F_{2 \mu \nu} = \partial _{\mu} F_{2 \nu} - \partial _{\nu} F_{2 \mu} 
+ i g {\rm tg} \alpha \lbrack F_{2 \mu} ~~,~~    F_{2 \nu} \rbrack ,
\eqno{(2.10)} 
$$
$$
B_{m \mu \nu} = \partial _{\mu} B_{m \nu} - \partial _{\nu} B_{m \mu} 
 ~~~,~~~~(m=1,2) .
\eqno{(2.11)} 
$$
Field strengths $F_{1 \mu \nu}$ and $F_{2 \mu nu}$ can be expressed as 
linear combinations of  
generators. That is:
$$
F_{m \mu \nu}= F^i_{m \mu \nu} \frac{\tau_i}{2}
~~,~~~~(m=1,2)
\eqno{(2.12)} 
$$
where $F^i_{m \mu \nu}$ are component field strengths whose expressions 
are:
$$
F_{1 \mu \nu}^i = \partial _{\mu} F_{1 \nu}^i - \partial _{\nu} F_{1 \mu}^i
+g \epsilon _{ijk} F_{1 \mu}^j    F_{1 \nu}^k
\eqno{(2.13)} 
$$
$$
F_{2 \mu \nu}^i = \partial _{\mu} F_{2 \nu}^i - \partial _{\nu} F_{2 \mu}^i
- g {\rm tg}\alpha \, \epsilon _{ijk} F_{2 \mu}^j    F_{2 \nu}^k
\eqno{(2.14)} 
$$
\\

~~~~ In order to introduce symmetry breaking, we will introduce  a scalar 
potential $v$ which we name it 
vacuum potential for the moment. It has mass dimension. It has no kinematic 
energy term in the 
lagrangian. So, it has no dynamical degree of freedom. The coupling between 
vacuum potential and 
matter fields can be regard as a kind of interactions between vacuum and 
matter fields. Because 
every Fock space  or every symmetry space has its own vacuum, we could 
select a vacuum 
potential 
for every particle or every symmetry. In this case, the ordinary mass term in 
the lagrangian can be  
rewritten as:
$$
-\frac{1}{2}  v^{\dag } \phi (x) \phi(x) v ~~,~~ - \overline{\psi}(x)  v  
\psi (x)
\eqno{(2.15)} 
$$
Because vacuum potential has no dynamical degree of freedom, it is hard to 
change its value. In the 
real physical world which is in a special phase of vacuum, its value in space-
time is even. In other 
words,  $v$ is always a constant in our world. Vacuum in different space has
different value. For $U(1)$ case, eq(2.15) changes into:
$$
-\frac{m ^2}{2} \phi (x) \phi(x) ~~,~~ -m \overline{\psi}(x) \psi (x),
\eqno{(2.16)} 
$$
where $m$ is the value of $v$. In the electroweak model, when $v$ has
definite value, the symmetry of the model will be broken, and quarks, leptons 
and gauge bosons will 
obtain masses.
\\

~~~~ The lagrangian density of the model is :
$$
{\cal L} = {\cal L} _l + {\cal L} _g + {\cal L} _{v-l} ,
\eqno{(2.17)} 
$$
where ${\cal L} _l , {\cal L} _g$ and ${\cal L} _{v-l}$ are the lagrangian 
density for leptons, 
lagrangian density for gauge fields and interaction lagrangian  between 
vacuum potential and leptons respectively. Their definitions respectively are:
$$
{\cal L }_l= - \overline{\psi}_L \gamma ^{\mu} 
(\partial _{\mu}+ \frac{i}{2}g \prime B_{1 \mu} -ig F_{1 \mu} ) \psi _L
- \overline{e}_R \gamma ^{\mu} 
(\partial _{\mu}+ ig \prime B_{1 \mu} ) e_R 
\eqno{(2.18)} 
$$
$$
\begin{array}{ccl}
{\cal L}_g &= &-\frac{1}{4}  F^{i \mu \nu}_1 F^i_{1 \mu \nu} 
- \frac{1}{4} F^{i \mu \nu}_2 F^i_{2 \mu \nu} 
-\frac{1}{4}  B^{\mu \nu}_1 B_{1 \mu \nu}  
-\frac{1}{4}  B^{\mu \nu}_2 B_{2 \mu \nu}  \\
&& - v^{\dag} 
\left \lbrack  {\rm cos} \theta _W ( {\rm cos} \alpha F_1^{\mu}+{\rm 
sin}\alpha F_2^{\mu}) -
{\rm sin}\theta _W ( {\rm cos}\alpha B_1^{\mu}+{\rm sin}\alpha 
B_2^{\mu} ) \right \rbrack \\
&&~\cdot 
\left \lbrack  {\rm cos} \theta _W ( {\rm cos} \alpha F_{1 \mu}+{\rm 
sin}\alpha F_{2 \mu}) -{\rm 
sin}\theta _W ( {\rm cos}\alpha B_{1 \mu}+{\rm sin}\alpha B_{2 \mu} ) 
\right \rbrack
v
\end{array}
\eqno{(2.19)} 
$$
$$
{\cal L} _{v-l} = -f (\overline{e}_R v^{\dag} \psi _L +\overline{\psi}_L v 
e_R) ,
\eqno{(2.20)} 
$$
where $f$ is a dimensionless parameter , $\alpha$ is a constant, $g, ~g'$ are 
coupling constants and 
$\theta_W$ are Weinberg angle. The relation of coupling constants and 
Weinberg angle is given by:
$$
{\rm tg} \theta_W = g' / g
\eqno{(2.21)} 
$$

~~~~ We have noticed that ${\cal L}_l$ and ${\cal L}_{v-l}$ are
completely the same as the corresponding parts of the standard model. And
there is no kinematic energy term for the vacuum potential $v$ in the
lagrangian. So $v$ is not a dynamical field.   
 \\

~~~~ Now, let's consider the symmetry of the model. The local $SU(2)_L$ 
gauge transformations of fields are:
$$
\psi _L \longrightarrow  U \psi_L
\eqno{(2.22)} 
$$
$$
e_R \longrightarrow  e_R
\eqno{(2.23)} 
$$
$$
F_{1 \mu} \longrightarrow  UF_{1 \mu} U^{\dag}
-\frac{1}{ig} U \partial_{\mu} U^{\dag}
\eqno{(2.24)} 
$$
$$
F_{2 \mu} \longrightarrow  UF_{2 \mu} U^{\dag}
+ \frac{1}{ig {\rm tg}\alpha } U \partial_{\mu} U^{\dag}
\eqno{(2.25)} 
$$
$$
B_{i \mu} \longrightarrow  B_{i \mu}
~~~~~(i=1,2)
\eqno{(2.26)} 
$$
$$
v  \longrightarrow  U v ,
\eqno{(2.27)} 
$$
where $U$ is the operator of local $SU(2)_L$ gauge transformation. The 
local $U(1)_Y$ gauge transformations of fields are:
$$
\psi _L \longrightarrow  e^{i \beta (x)} \psi_L
\eqno{(2.28)} 
$$
$$
e_R \longrightarrow  e^{2 i \beta (x)} e_R
\eqno{(2.29)} 
$$
$$
F_{i \mu} \longrightarrow  F_{i \mu}
~~~~(i=1,2)
\eqno{(2.30)} 
$$
$$
B_{1 \mu} \longrightarrow  B_{1 \mu}
- \frac{2}{g' } \partial_{\mu} \beta (x)
\eqno{(2.31)} 
$$
$$
B_{2 \mu} \longrightarrow  B_{2 \mu}
+ \frac{2}{g' {\rm tg} \alpha} \partial_{\mu} \beta (x)
\eqno{(2.32)} 
$$
$$
v  \longrightarrow e^{ - i \beta (x)} v .
\eqno{(2.33)} 
$$
It is easy to prove that the lagrangian defined by eq(2.17-20) is invariant 
under the above local 
$SU(2)_L \times U(1)_Y$ gauge transformations, so the lagrangian has 
strict local $SU(2)_L \times U(1)_Y$ gauge symmetry.  \\

~~~~ In the above lagrangian, there are two sets of gauge fields. The second 
set of gauge fields is the compensatory fields of the first set. In other words, 
the gauge transformations of 
the first set of gauge fields are determined by the gauge transformation of 
matter fields, and the 
gauge transformations of the second set of gauge fields are determined by the 
transformation the first set of 
gauge fields. So, there is no restriction on gauge transformations and the 
model has the maximal 
local $SU(2)_L \times U(1)_Y$ gauge symmetry. In the same time, we 
should notice that $F_{1 
\mu},~ F_{2 \mu},~B_{1 \mu}$ and $B_{2 \mu}$ are all standard gauge 
fields. In the present case, 
we do not let gauge fields $F_{2 \mu}$ and $B_{2 \mu}$ interact with 
matter fields. According to 
reference \lbrack 8 \rbrack, after introducing a new parameter, we could let 
both two sets of gauge 
fields interact directly with matter fields. But we do not do so in this paper so 
as to keep gauge fields 
minimal couple to
matter fields in the original lagrangian. 
\\

\section{Symmetry breaking and masses of particles }

~~~~ We have already said that $v$ represents the influence of vacuum 
above. Although in the original lagrangian density, $v$ has the degree of 
freedom of gauge transformation, in our real physical world, the state of 
vacuum can not be varied freely and the  properties of vacuum are rather 
stable, it has no gauge transformation degree of freedom. In  the local 
inertial coordinate system,  vacuum is invariant under space-time  
translation. So $v$ is well-distributed, it is a constant. Suppose that $v$ has 
the following value
$$
v =\left ( 
\begin{array}{c}
{\rm v}_1\\
{\rm v}_2
\end{array}
\right ),
\eqno{(3.1)} 
$$
where ${\rm v}_1$ and ${\rm v}_2$ satisfy the following relation:
$$
{\rm v}_1^2 + {\rm v}_2^2 = \mu^2 /2 ,
\eqno{(3.2)} 
$$
where $\mu$ is a constant with mass dimension. Please note that eq(3.1) is 
the most general form of two-component vector constant $v$. But it is not the 
correct form of the 
vacuum of our world. So we make a global $SU(2)_L$ gauge transformation 
so as to make $v$ 
change into the following form
$$
v =\left ( 
\begin{array}{c}
0\\
\mu / \sqrt{2}
\end{array}
\right ).
\eqno{(3.3)} 
$$
Because the lagrangian has global $SU(2)_L \times U(1)_Y$ gauge 
symmetry, its form keeps 
unchanged under the above global transformation. It is known that the
properties of vacuum affect the dynamical properties of our physical world.
In present case, different forms of vacuum potential $v$ will give different
forms of lagrangian after symmetry breaking. In our real physical world, the
electroweak vacuum potential is given by eq(3.3). We could understand 
eq(3.3) from another point 
of 
view. Although in the original lagrangian vacuum has very high symmetry, in 
our real physical 
world, 
it does not have so high symmetry. The vacuum of our real world is stable, so 
it is in a special gauge 
which is determined by eq(3.3). We adopt eq(3.3) means that we have select 
a special gauge. So, 
eq(3.3) is the gauge condition of our world. When we select a special gauge, 
the symmetry of the 
model is broken at the same time. So, in this point of view,  symmetry 
breaking is originated from gauge fixing of the vacuum. \\

~~~~ After symmetry breaking, gauge fields $F_{1 \mu} , ~ F_{2 \mu},~ 
B_{1 \mu}$ and $B_{2 
\mu}$ are not eigenvectors of mass matrix, so they don't correspond to the 
fields of particles in the real physical world. In order to obtain eigenvectors of 
mass matrix,  we will  make the following two sets of 
transformations of fields. The first set of transformations are:
$$
W_{\mu}={\rm cos}\alpha F_{1 \mu}+{\rm sin}\alpha F_{2 \mu}
\eqno{(3.4)} 
$$
$$
W_{2 \mu}=-{\rm sin}\alpha F_{1 \mu}+{\rm cos}\alpha F_{2 \mu}
\eqno{(3.5)} 
$$
$$
C_{1 \mu}={\rm cos}\alpha B_{1 \mu}+{\rm sin}\alpha B_{2 \mu}
\eqno{(3.6)} 
$$
$$
C_{2 \mu}=-{\rm sin}\alpha B_{1 \mu}+{\rm cos}\alpha B_{2 \mu} .
\eqno{(3.7)} 
$$
The second set of transformations are:
$$
Z_{\mu}= {\rm sin}\theta _W C_{1 \mu}-{\rm cos}\theta _W W^3_{ \mu}
\eqno{(3.8)} 
$$
$$
A_{\mu}= {\rm cos}\theta _W C_{1 \mu}+{\rm sin}\theta _W W^3_{ \mu}
\eqno{(3.9)} 
$$
$$
Z_{2 \mu}= {\rm sin}\theta _W C_{2 \mu}-{\rm cos}\theta _W W^3_{2 
\mu}
\eqno{(3.10)} 
$$
$$
A_{2 \mu}= {\rm cos}\theta _W C_{2 \mu}+{\rm sin}\theta _W W^3_{2 
\mu}. 
\eqno{(3.11)} 
$$
The first set of transformations are the standard transformations discussed in
references \lbrack 7 \rbrack and \lbrack 8 \rbrack. The second set of
transformations are the standard transformations used in the standard model.
After these two sets of field transformations, all terms of  two-body 
coupling of gauge fields disappear from the lagrangian. \\

~~~~ As we have mentioned at the end of chapter two, fields $F_{1 \mu},~
F_{2 \mu}, ~B_{1 \mu}$ and $B_{2 \mu}$ are all standard gauge fields.
Although, after the first set of field transformation, the gauge
transformations of $W_{\mu}$ and $C_{1 \mu}$ are not in the standard 
forms
of gauge transformation, for the sake of convenience, we call them general
gauge fields, or simply call them gauge fields. It is obvious that $W_{\mu}$
and $C_{1 \mu}$ are not ordinary vector fields, because they are "made 
from"
gauge fields (according to eq(3.4) and eq(3.6)) and they transmit
interactions between matter fields. Therefore, it is not suitable to call
them ordinary vector fields or to regard them as ordinary matter fields. We 
call them general gauge fields for the present.
Similarly, we call $W_{\mu}^{\pm},~ Z_{\mu}$ and $A_{\mu}$ general 
gauge
fields. \\

~~~~ After all these transformations, the lagrangian densities of the model 
change into:
$$
\begin{array}{ccl}
{\cal L}_l +{\cal L}_{v-l} &= & - \overline{e} (\gamma ^{\mu} 
\partial _{\mu}+ \frac{1}{\sqrt{2}} f \mu ) e 
-\overline{\nu}_L \gamma ^{\mu} \partial _{\mu}\nu _L\\
&&
+\frac{1}{2} \sqrt{g^2 + {g \prime}^2} {\rm sin}2\theta_W  
j^{em}_{\mu} 
 ( {\rm cos}\alpha A^{\mu}- {\rm sin}\alpha  A^{\mu}_2  )  \\
&&
- \sqrt{g^2 + {g \prime}^2} j^{z}_{\mu} 
( {\rm cos}\alpha Z^{\mu} - {\rm sin}\alpha Z_2^{ \mu} ) \\
&&
+ \frac{\sqrt{2}}{2} ig \overline{\nu}_L \gamma ^{\mu} e_L 
( {\rm cos}\alpha W_{\mu}^{+} - {\rm sin}\alpha W_{2 \mu}^{+} )  \\
&&
+ \frac{\sqrt{2}}{2} ig \overline{e}_L \gamma ^{\mu} {\nu}_L 
( {\rm cos}\alpha W_{\mu}^{-} - {\rm sin}\alpha W_{2 \mu}^{-} )
\end{array}
\eqno{(3.12)} 
$$
$$
\begin{array}{ccl}
{\cal L}_g &= &-\frac{1}{2}  W^{+ \mu \nu}_0  W^{-}_{0 \mu \nu} 
-\frac{1}{4}  Z^{\mu \nu} Z_{ \mu \nu} 
-\frac{1}{4}  A^{\mu \nu} A_{ \mu \nu} 
 \\
&&-\frac{1}{2}  W^{+ \mu \nu}_{2 0}  W^{-}_{2 0 \mu \nu} 
-\frac{1}{4}  Z^{\mu \nu}_2  Z_{2  \mu \nu} 
-\frac{1}{4}  A^{\mu \nu}_2  A_{2  \mu \nu} 
\\
&& -\frac{\mu ^2}{2}  Z^{\mu }  Z_{ \mu } 
-\mu ^2 {\rm cos}^2 \theta _W  W^{+ \mu}  W^{-}_{\mu} +{\cal L}_{g 
I}
\end{array} ,
\eqno{(3.13)} 
$$
where ${\cal L}_{g I}$ only contains interaction terms of gauge fields, and 
$$
W^{\pm}_{m \mu } = \frac{1}{\sqrt{2}} (W^1_{m \mu} \mp i W^2_{m 
\mu} )
~~~( m=1,2, ~W_{1 \mu} \equiv W_{\mu}  ) .
\eqno{(3.14)} 
$$
In the above lagrangian, the field strengths of gauge fields are defined as:
$$
W^{\pm}_{m0 \mu \nu} = \partial _{\mu} W^{\pm}_{m \nu} 
- \partial _{\nu} W^{\pm}_{m \mu} 
~~~( m=1,2, ~W^{\pm}_{1 \mu} \equiv W^{\pm}_{\mu}  ) ,
\eqno{(3.15)} 
$$
$$
Z_{m \mu \nu} = \partial _{\mu} Z_{m \nu} 
- \partial _{\nu} Z_{m \mu} 
~~~( m=1,2, ~Z_{1 \mu} \equiv Z_{\mu}  ) ,
\eqno{(3.16)} 
$$
$$
A_{m \mu \nu} = \partial _{\mu} A_{m \nu} 
- \partial _{\nu} A_{m \mu} 
~~~( m=1,2, ~A_{1 \mu} \equiv A_{\mu}  ) .
\eqno{(3.17)} 
$$
The currents in the above lagrangian are defined as:
$$
j_{ \mu }^{em} = -i \overline{e} \gamma_{\mu} e
\eqno{(3.18)} 
$$
$$
j_{ \mu }^{Z} =j_{\mu}^{3} - {\rm sin}^2 \theta_W  j_{\mu}^{em}
= i \overline{\psi}_L  \gamma_{\mu} \frac{\tau ^3}{2} \psi_L
- {\rm sin}^2 \theta_W  j_{\mu}^{em} .
\eqno{(3.19)} 
$$
\\

~~~~ From the above lagrangian, we could see that the mass of fermion $e$ 
is $\frac{1}{\sqrt{2}} f 
\mu$, the mass of neutrino is zero, the masses of charged intermediate gauge 
bosons $W^{\pm}$ are 
$\mu {\rm cos} \theta_W$, the mass of neutral intermediate gauge boson $Z$ 
is $\mu = 
\frac{m_W}{{\rm cos} \theta_W}$ and all other gauge fields are massless. 
That is
$$
m_e =  \frac{1}{\sqrt{2}} f \mu~~~,~~~
m_{\nu}=0
\eqno{(3.20)} 
$$
$$
m_W =  \mu {\rm cos} \theta_W ~~~,~~~
m_Z = \mu = \frac{m_W}{{\rm cos} \theta_W}
\eqno{(3.21)} 
$$
$$
m_A= m_{A2}=m_{W2}=m_{Z2}=0
\eqno{(3.22)} 
$$
It is easy to see that, in this model, the expressions of the masses of fermions 
and intermediate gauge 
bosons are the same as those in the  standard model. \\

\section{Compare with the Standard Model }

~~~~ We have known that the standard  model is a successful model in 
describing electroweak interactions when the energy of the system is not very 
high. If the new electroweak model is also a successful model in describing 
high energy electroweak interactions, it should be able to return to the 
standard model in a limit. The lagrangian of the new electroweak  model 
discussed
above is quite different in form from that of the standard model. And  all
these differences concern of the parameter $\alpha$ and Higgs field.
Parameter $\alpha$ is a free parameter whose value is not determined at
present. If we let the value of $\alpha$ vary, then in a proper limit, 
the model discussed above  will approximately return to the standard model. 
Now, let's discuss the standard model limit of the new electroweak model. 
Suppose that  parameter $\alpha$ is much smaller than 1,
$$
\alpha \ll 1 ,
\eqno{(4.1)} 
$$
then in the leading term approximation,
$$
{\rm cos}\alpha \approx 1~~~,~~~
{\rm sin}\alpha \approx 0 .
\eqno{(4.2)} 
$$
In this case, the lagrangian density for fermions becomes:
$$
\begin{array}{ccl}
{\cal L}_l +{\cal L}_{v-l} &= & - \overline{e} (\gamma ^{\mu} 
\partial _{\mu}+ \frac{1}{\sqrt{2}} f \mu ) e 
-\overline{\nu}_L \gamma ^{\mu} \partial _{\mu}\nu _L\\
&&
+{\rm e}  j^{em}_{\mu} A^{\mu}- \sqrt{g^2 + {g \prime}^2} 
j^{z}_{\mu} 
Z^{\mu}\\
&&
+ \frac{\sqrt{2}}{2} ig \overline{\nu}_L \gamma ^{\mu} e_L 
W_{\mu}^{+}
+ \frac{\sqrt{2}}{2} ig \overline{e}_L \gamma ^{\mu} {\nu}_L 
W_{\mu}^{-}
\end{array}
\eqno{(4.3)} 
$$
where 
$$
{\rm e} = \frac{g  g'}{\sqrt{g^2 + {g' }^2}} 
\eqno{(4.4)} 
$$
is the coupling constant of electromagnetic interactions. From eq(4.3), 
we see ${\cal L}_l +{\cal L}_{v-l}$ is the same as the corresponding 
lagrangian density in the standard model. Besides, in this approximation, the 
form of ${\cal L}_{g I}$ is simplified and the terms in ${\cal L}_{g }$ other 
than ${\cal L}_{gI}$ do not change. Therefore, in this approximation, except 
for the terms concern Higgs particle, the 
lagrangian of the model discussed in this paper is almost the same as that of 
the standard model: they 
have the same mass relation of the intermediate gauge bosons, the same 
charged currents and neutral 
current, the same electromagnetic current, the same coupling constant of the 
electromagnetic 
interactions, the same effective coupling constant of weak interactions
$\cdots$ etc.. On the other 
hand, we must see that there are two fundamental differences between the 
new electroweak model 
and the standard model: 1) there is no Higgs particle in the new electroweak 
model, so there are no 
interaction terms between Higgs particle and leptons, quarks or 
gauge bosons; 2)compare with the standard 
model, we have introduced two sets of gauge fields in the new electroweak 
model. These new gauge 
bosons are all massless. In the limit $\alpha \longrightarrow 0$, the coupling 
between massless gauge bosons and quarks or leptons will also go to zero. 
Because
$\alpha$ is very small 
(which we will discussed later), the influences of these massless gauge 
bosons to the course of electroweak 
interactions will be small too. No differences between these two electroweak
theories are detected by experiments. Therefore, we could anticipate that, if
parameter $\alpha$ is small, two electroweak theories will give similar
results in describing  low energy electroweak interactions of quarks and leptons,
and there will exist no contradictions between new electroweak model and
experiments on electroweak interactions. 
\\

~~~~ In the present model, we have introduced eight kinds of gauge bosons, 
they are: $W^{\pm}, ~Z, 
~A, ~W_2^{\pm}, ~Z_2 $ and $A_2$. Those four kinds of gauge bosons in 
front have already been  introduced in 
the standard model, others are introduced by the present model. So, in the 
present theory, there are 
two kinds of charged massive gauge bosons, one kind of neutral massive 
gauge boson, three kinds of 
neutral massless gauge bosons and two kinds of charged massless gauge
bosons. \\

~~~~ There are two different kinds of  intermediate gauge bosons which 
couple to matter fields in different manners. The coupling 
constants between leptons and massive charged intermediate gauge bosons or 
massless charged 
intermediate gauge bosons respectively are:
$$
g~{\rm cos}\alpha ~~~,~~~ g~{\rm sin}\alpha
\eqno{(4.5)} 
$$
There are also two different coupling constants between leptons and massive 
neutral intermediate gauge boson or massless neutral intermediate gauge 
bosons, they are 
$$
\sqrt{g^2 + {g \prime}^2} {\rm cos}\alpha ~~~,~~~ \sqrt{g^2 + {g 
\prime}^2} {\rm sin}\alpha
\eqno{(4.6)} 
$$
The effective coupling constant of Fermi weak interactions caused by 
massive intermediate gauge 
bosons is 
$$
\frac{G}{\sqrt{2}} = \frac{g^2}{8 m_W^2}{\rm cos}^2 \alpha .
\eqno{(4.7)} 
$$
When the parameter $\alpha$ is very small, the above relation will return to 
the well-known relation 
in the standard model. That means that, in the low-energy phenomena, the
new electroweak theory will give a correct description on Fermi weak
interactions. \\

~~~~ There exist two different kinds of electromagnetic fields $A_{\mu}$ 
and $A_{2 \mu}$, 
correspondingly, there exist two different kinds coupling constants of 
electromagnetic interactions. 
They respectively are:
$$
{\rm e} _1 = \frac{g  g'}{\sqrt{g^2 + {g' }^2}} {\rm cos} \alpha 
= {\rm e}  ~  {\rm cos} \alpha
\eqno{(4.8)} 
$$
$$
{\rm e} _1 = \frac{g  g'}{\sqrt{g^2 + {g' }^2}} {\rm sin} \alpha 
= {\rm e}  ~  {\rm sin} \alpha
\eqno{(4.9)} 
$$
The electromagnetic field exist in nature should be a mixture of the two 
different kinds 
electromagnetic fields $A_{\mu}$ and $A_{2 \mu}$. In other words, the 
electromagnetic 
interactions in nature should be transmitted by both $A_{\mu}$ and $A_{2 
\mu}$. The effective 
coupling constant of electromagnetic interactions should be: 
$$
{\rm e} ^2={\rm e} _1^2+{\rm e} _2^2 ~~~,~~~
{\rm e}  = \frac{g  g'}{\sqrt{g^2 + {g' }^2}}
\eqno{(4.10)} 
$$
So, the effective coupling constant of electromagnetic interactions is the same 
as the coupling 
constant of electromagnetic interactions in the standard model. Furthermore, 
the value of the 
parameter $\alpha$ does not affect the value of effective coupling constant of 
electromagnetic 
interactions. Therefore, although we have introduced two different kinds of 
electromagnetic fields, 
the law for electromagnetic interactions will not changed.  And, no matter
how much is the parameter $\alpha$, two electroweak theories will give the
same results  in describing pure electromagnetic interaction course. \\

~~~~ There are three kinds of massless neutral gauge bosons in the new 
theory. they are $Z_2,~A$ 
and $A_2$. They are all stable particles and  interact with quarks or 
leptons. In other words, they 
have similar interaction properties. The most important phenomenological 
difference between photon 
and $Z_2$ particle is that $Z_2$ particle interacts with neutrino but photon 
does not interact with 
neutrino. But this difference is hard to detect in the experiment. In other 
words, if there is  $Z_2$ 
particle mixed in photon, it is hard to distinguish them.  If physicists found 
that photon take part in weak interactions 
much stronger than expected, that means that there is likely $Z_2$ particle 
mixed in photon. Some more discussions on them will be presented in the
final chapter. \\

~~~~ It is  known that there are two kinds of long-range force fields:
gravitation field and electromagnetic field. If $Z_2$ particle exists
in nature, there will be a new kind of long-range force
field transmitted by massless $Z_2$ particle. Some more discussions on this
problem can be found at the end of this paper. \\

~~~~ In a word, when parameter $\alpha$ is small enough, the new 
electroweak
model will approximately return to the standard model. Because the
theoretical predictions of the standard model coincide well with
experimental results, we could believe that the parameter $\alpha$ will be
very small. But, even if $\alpha \longrightarrow 0$, the new electroweak
model can not completely return to the standard model, because there exists
no Higgs particle in the new electroweak model and there exist no 
$W_2^{\pm}$
and $Z_2$ particles in the standard model. Up to now, no differences 
between
these two electroweak model are detected by experiments, so, at present, it is
hard to say that which model is the correct model in describing electroweak
interactions. Some more discussions on these two models will be presented at
the end of this paper. \\

\section{ Electroweak interactions of quarks }

~~~~ Now, let's discuss the electroweak interactions of quarks. It is  known
that, up to now, there are three generations of quarks which have six different 
flavours. Quarks take part in strong interactions, electromagnetic interactions 
and weak interactions. In this section, we will use the general gauge field 
theory \lbrack 7-8 \rbrack to construct the electroweak model for quarks. 
\\

~~~~ According to the standard model, there exists mixing between three 
different 
kinds of quarks $d,s$ and $b$ \lbrack 10 \rbrack. Define:
$$
\left (
\begin{array}{c}
d_{\theta} \\
s_{\theta}  \\
b_{\theta}
\end{array}
\right )
= K 
\left (
\begin{array}{c}
d\\
s\\
b
\end{array} 
\right ),
\eqno{(5.1)} 
$$
where $K$ is the Kabayashi-Maskawa mixing matrix whose general form is:
$$
K = 
\left (
\begin{array}{ccc}
c_1 & s_1 c_3 & s_1 s_3 \\
-s_1 c_2 & c_1 c_2 c_3 - s_2 s_3 e^{i \delta} 
& c_1 c_2 s_3 + s_2 c_3 e^{i \delta}   \\
s_1 s_2 & -c_1 s_2 c_3 -c_2 s_3 e^{i \delta} 
& -c_1 s_2 s_3 +c_2 c_3 e^{i \delta}
\end{array}
\right )
\eqno{(5.2)} 
$$
where 
$$
c_i = {\rm cos} \theta_i ~~,~~~~ s_i = {\rm sin} \theta_i ~~~(i=1,2,3)
\eqno{(5.3)} 
$$
and $\theta_i$ are generalized Cabibbo angles. \\

~~~~ According to the standard model, quarks form left-hand doublets and 
right-hand singlets. Denote:
$$
q_L^{(1)} = 
\left (
\begin{array}{c}
u_L \\
d_{\theta L}
\end{array}
\right ) ,~~
q_L^{(2)} = 
\left (
\begin{array}{c}
c_L \\
s_{\theta L}
\end{array} 
\right ),~~
q_L^{(3)} = 
\left (
\begin{array}{c}
t_L \\
b_{\theta L}
\end{array}
\right ) 
\eqno{(5.4)} 
$$
$$
\begin{array}{ccc}
U_R^{(1)}= u_R
& U_R^{(2)}= c_R
& U_R^{(3)}= t_R \\
D_{ \theta R}^{(1)}= d_{\theta R}
& D_{ \theta R}^{(2)}= s_{\theta R}
& D_{ \theta R}^{(3)}= b_{\theta R}
\end{array}
\eqno{(5.5)} 
$$
It is known that left-hand doublets have weak isospin $\frac{1}{2}$ and 
weak 
hypercharge  $\frac{1}{3}$, right-hand singlets  have no weak isospin, 
$U_R^{(j)}$s have weak hypercharge $\frac{4}{3}$ and $D_{\theta 
R}^{(j)}$s 
have weak hypercharge $ - \frac{2}{3}$. \\

~~~~ The lagrangian of the model consists of three parts:
$$
{\cal L} = {\cal L} _q + {\cal L} _g + {\cal L} _{v-q}
\eqno{(5.6)}
$$
where ${\cal L} _q$ is the lagrangian density of quark fields, ${\cal L} _g$
is the lagrangian density of gauge fields which is given by eq.(2.19) and
${\cal L} _{v-q}$ contains the interaction terms among quarks fields and 
vacuum
potential. The lagrangian density of quark fields is:
$$
\begin{array}{ccl}
{\cal L}_q &= &-\sum_{j=1}^{3} \overline{q}_L^{(j)} \gamma^{\mu}
( \partial_{\mu}-ig F_{1 \mu}- \frac{i}{6}g' B_{1 \mu}) q_L^{(j)}
-\sum_{j=1}^{3} \overline{U}_R^{(j)} \gamma^{\mu} 
( \partial_{\mu}-i \frac{2}{3} g' B_{1\mu} ) U_R^{(j)}  \\
&& -\sum_{j=1}^{3} \overline{D}_{\theta R}^{(j)} \gamma^{\mu} 
( \partial_{\mu} + i \frac{1}{3} g' B_{1\mu} ) D_{\theta R}^{(j)}  
\end{array}
\eqno{(5.7)}
$$
And the interaction lagrangian density of quark fields and vacuum
potential is:
$$
{\cal L}_{v-q } = -\sum_{j=1}^{3} ( f^{(j)} \overline{q}_L^{(j)} 
\overline{v}
U_R^{(j)} + f^{(j) \ast} \overline{U}_R^{(j)} \overline{v}^{\dag} 
q_L^{(j)} )
-\sum_{j,k=1}^{3} ( f^{(jk)} \overline{q}_L^{(j)} v D_{\theta R}^{(k)} 
+ f^{(jk) \ast} \overline{D}_{\theta R}^{(k)} v^{\dag} q_L^{(j)} )
\eqno{(5.8)}
$$
where
$$
\overline{v} = i \sigma_{2} v^{\ast} =
\left (
\begin{array}{c}
v_2^{\dag} \\
- v_1^{\dag}
\end{array}
\right )
\eqno{(5.9)}                                              
$$    

~~~~ The lagrangian density given by eq(5.6-8) is invariant under the
following local $SU(2)_L$ gauge transformation:
$$
q_L^{(j)} \longrightarrow U q_L^{(j)}
\eqno{(5.10)}
$$
$$
U_R^{(j)} \longrightarrow U_R^{(j)}
\eqno{(5.11)}
$$
$$
D_{\theta R}^{(j)} \longrightarrow D_{\theta R}^{(j)}
\eqno{(5.12)}
$$
$$
F_{1 \mu} \longrightarrow U F_{1 \mu} U^{\dag} - \frac{1}{i g} U
\partial_{\mu} U^{\dag}
\eqno{(5.13)}
$$
$$
F_{2 \mu} \longrightarrow U F_{2 \mu} U^{\dag} + \frac{1}{i g {\rm tg}
\alpha} U \partial_{\mu} U^{\dag}
\eqno{(5.14)}
$$
$$
B_{m \mu} \longrightarrow B_{m \mu}~~~~~(m=1,2)
\eqno{(5.15)}
$$
$$                                                              
v \longrightarrow U v
\eqno{(5.16)}
$$
$$                                                              
\overline{v} \longrightarrow U \overline{v}
\eqno{(5.17)}
$$
and the following local $U(1)_Y$ gauge transformations:
$$
q_L^{(j)} \longrightarrow e^{-i \beta /3} q_L^{(j)}
\eqno{(5.18)}
$$
$$
U_R^{(j)} \longrightarrow e^{-4i \beta /3} U_R^{(j)}
\eqno{(5.19)}
$$
$$
D_{\theta R}^{(j)} \longrightarrow e^{2i \beta /3}D_{\theta R}^{(j)}
\eqno{(5.20)}
$$
$$
F_{m \mu} \longrightarrow F_{m \mu}~~~~~(m=1,2)
\eqno{(5.21)}
$$
$$
B_{1 \mu} \longrightarrow B_{1 \mu} - \frac{2}{g'} \partial_{\mu} \beta
\eqno{(5.22)}
$$
$$                                             
B_{2 \mu} \longrightarrow B_{2 \mu} 
+ \frac{2}{g'{\rm tg} \alpha} \partial_{\mu} \beta
\eqno{(5.23)}
$$
$$
v \longrightarrow e^{-i \beta} v
\eqno{(5.24)}
$$
$$                                                  
\overline{v} \longrightarrow e^{i \beta} \overline{v}.        
\eqno{(5.25)}
$$
In other words, the lagrangian density define by eq.(5.6) has strict local
$SU(2)_L \times U(1)_Y$ gauge symmetry. \\

~~~~ After symmetry breaking, vacuum potential $v$ has the form of 
eq.(3.3).
Correspondingly, the form of $\overline{v}$ is:
$$
\overline{v} = 
\left (
\begin{array}{c}
\frac{\mu}{\sqrt{2}} \\
0
\end{array}
\right ).
\eqno{(5.26)}
$$
Then, we make a sequence of transformations of gauge fields defined by
eq.(3.3-11). After these transformations of gauge fields, the lagrangian
density of gauge fields becomes the form given by eq.(3.13).
Correspondingly, the lagrangian density of quark fields becomes:
$$
\begin{array}{ccl}
{\cal L}_q &= & -\overline{u} \partial \!\!\! / \, u 
 -\overline{d} \partial \!\!\! / \, d
 -\overline{c} \partial \!\!\! / \, c
 -\overline{s} \partial \!\!\! / \, s
 -\overline{t} \partial \!\!\! / \, t
 -\overline{b} \partial \!\!\! / \, b  \\
&&
+\frac{1}{2} \sqrt{g^2 + {g \prime}^2} {\rm sin}2\theta_W  
j^{em}_{\mu} 
 ( {\rm cos}\alpha A^{\mu}- {\rm sin}\alpha  A^{\mu}_2  )  \\
&&
- \sqrt{g^2 + {g \prime}^2} j^{z}_{\mu} 
( {\rm cos}\alpha Z^{\mu} - {\rm sin}\alpha Z_2^{ \mu} ) \\
&&
+ \frac{\sqrt{2}}{2} ig (\overline{u}_L \gamma ^{\mu} d_{\theta L}
+\overline{c}_L \gamma ^{\mu} s_{\theta L}
+\overline{t}_L \gamma ^{\mu} b_{\theta L}) 
( {\rm cos}\alpha W_{\mu}^{+} - {\rm sin}\alpha W_{2 \mu}^{+} )  \\
&&
+ \frac{\sqrt{2}}{2} ig (\overline{d}_{\theta L} \gamma ^{\mu} u_L
+\overline{s}_{\theta L} \gamma ^{\mu} c_L
+\overline{b}_{\theta L} \gamma ^{\mu} t_L )
( {\rm cos}\alpha W_{\mu}^{-} - {\rm sin}\alpha W_{2 \mu}^{-} )
\end{array}
\eqno{(5.27)}
$$
In the above equation, currents are defined by the following relations:
$$
j_{ \mu }^{em} = i ( \frac{2}{3} \overline{u} \gamma_{\mu} u
- \frac{1}{3} \overline{d} \gamma_{\mu} d
+ \frac{2}{3} \overline{c} \gamma_{\mu} c 
- \frac{1}{3} \overline{s} \gamma_{\mu} s
+ \frac{2}{3} \overline{t} \gamma_{\mu} t 
- \frac{1}{3} \overline{b} \gamma_{\mu} b )
\eqno{(5.28)} 
$$
$$
j_{ \mu }^{Z} =j_{\mu}^{3} - {\rm sin}^2 \theta_W  j_{\mu}^{em}.
\eqno{(5.29)} 
$$
$$
\begin{array}{ccl}
j^3_{\mu} & = & \sum_{j=1}^3 i \overline{q}_L^{(j)} 
\gamma_{\mu} \frac{\tau^3}{2} q_L^{(j)}  \\
& = & \frac{i}{2}
( \overline{u}_L \gamma_{\mu} u_L
- \overline{d}_L \gamma_{\mu} d_L
+ \overline{c}_L \gamma_{\mu} c_L
- \overline{s}_L \gamma_{\mu} s_L
+ \overline{t}_L \gamma_{\mu} t_L
- \overline{b}_L \gamma_{\mu} b_L )
\end{array}
\eqno{(5.30)}
$$
And the lagrangian density for the interactions among quark fields and 
vacuum potential becomes:
$$
{\cal L}_{v-q } = - \frac{\mu}{\sqrt{2}}\sum_{j=1}^{3} 
( f^{(j)} \overline{u}_L^{(j)} U_R^{(j)} 
+ f^{(j) \ast} \overline{U}_R^{(j)}  u_L^{(j)} )
-\frac{\mu}{\sqrt{2}} \sum_{j,k=1}^{3} 
( f^{(jk)} \overline{d}_{\theta L}^{(j)} D_{\theta R}^{(k)}
+ f^{(jk) \ast} \overline{D}_{\theta R}^{(k)}  d_{\theta L}^{(j)} )
\eqno{(5.31)}
$$
\\

~~~~ Denote
$$
F=( f^{(jk)} )
\eqno{(5.32)}
$$
is a $3 \times 3$ matrix. In eq(5.8), parameters $f^{(j)}$ and $f^{(jk)}$ are
selected to satisfy the following requirements: 
$$
f^{(j) \ast} = f^{(j)} ~,~~~~f^{(jk) \ast} = f^{(kj)}.
\eqno{(5.33)}
$$
So, matrix $F$ is an Hermit matrix, it could be diagonalized through
similarity transformation. In electroweak model, matrix $F$ is selected 
to have the following form:
$$
F = \frac{\sqrt{2}}{\mu} K M_D K^{\dag}
\eqno{(5.34)}
$$
where $K$ is the similarity transformation matrix which is defined by
eq(5.2) and $M_D$ is a diagonal
matrix whose form is:
$$
M_D =
\left (
\begin{array}{ccc}
m_d &&  \\
& m_s &  \\
&& m_b
\end{array}
\right ).
\eqno{(5.35)}
$$
Because $K$ is a unitary matrix, lagrangian density ${\cal L}_{v-q}$ 
becomes
$$
{\cal L}_{v-q } = 
- m_u \overline{u} u - m_d \overline{d} d
- m_c \overline{c} c - m_s \overline{s} s
- m_t \overline{t} t - m_b \overline{b} b
\eqno{(5.36)}
$$
where,
$$
m_u = f^{(1)} \mu / \sqrt{2} ~,~~~
m_c = f^{(2)} \mu / \sqrt{2} ~,~~~
m_t = f^{(3)} \mu / \sqrt{2} .
\eqno{(5.37)}
$$
\\

~~~~ After all these operations, we could obtain the following result:
$$
\begin{array}{ccl}
{\cal L}_q + {\cal L}_{v-q} &= & 
-\overline{u} (\partial \!\!\! / \,+ m_u) u
-\overline{c} (\partial \!\!\! / \,+ m_c) c
-\overline{t} (\partial \!\!\! / \,+ m_t) t  \\
&&
-\overline{d} (\partial \!\!\! / \,+ m_d) d
-\overline{s} (\partial \!\!\! / \,+ m_s) s
-\overline{b} (\partial \!\!\! / \,+ m_b) b   \\
&&
+\frac{1}{2} \sqrt{g^2 + {g \prime}^2} {\rm sin}2\theta_W  
j^{em}_{\mu}
 ( {\rm cos}\alpha A^{\mu}- {\rm sin}\alpha  A^{\mu}_2  )  \\
&&
- \sqrt{g^2 + {g \prime}^2} j^{z}_{\mu}
( {\rm cos}\alpha Z^{\mu} - {\rm sin}\alpha Z_2^{ \mu} ) \\
&&
+ \frac{\sqrt{2}}{2} ig (\overline{u}_L \gamma ^{\mu} d_{\theta L}
+\overline{c}_L \gamma ^{\mu} s_{\theta L}
+\overline{t}_L \gamma ^{\mu} b_{\theta L})
( {\rm cos}\alpha W_{\mu}^{+} - {\rm sin}\alpha W_{2 \mu}^{+} )  \\
&&
+ \frac{\sqrt{2}}{2} ig (\overline{d}_{\theta L} \gamma ^{\mu} u_L
+\overline{s}_{\theta L} \gamma ^{\mu} c_L
+\overline{b}_{\theta L} \gamma ^{\mu} t_L )
( {\rm cos}\alpha W_{\mu}^{-} - {\rm sin}\alpha W_{2 \mu}^{-} )
\end{array}
\eqno{(5.38)}
$$
\\

~~~~ In the limit $\alpha \longrightarrow 0$, except for those terms concern
of Higgs field, the electroweak model for quarks discussed in this chapter
will also approximately return to the standard model. The new electroweak 
model and the
standard model have the same electromagnetic current, the same neutral
current, the same charged currents, the same coupling constants for weak 
and
electromagnetic interactions and the same expressions for quark masses. If
$\alpha$ is very small, the coupling between quarks and massless
intermediate gauge bosons is also very small. Therefore, the correction
cause by massless intermediate gauge bosons will be very small and 
the effects of interactions between quarks and massless intermediate gauge 
bosons are hard to be detected by experiments. In  a  words, except for
Higgs particle and interactions between Higgs particle and quarks, the new
electroweak model keeps almost all other dynamical properties of the standard
model. \\

\section{Discussions}

~~~~ The main goal of this paper is to construct an electroweak model in
which we avoid using Higgs mechanism and avoid introducing Higgs particle. 
We know that, up to now, although the energy of 
accelerated particles has already 
reached several Tev, experimental physicists don't find any evidence of the
existence of Higgs particle. Besides, the mass of Higgs particle predicted by 
theory has been rising from several Gev to several hundred Gev. This 
situation gives us an impression that Higgs particle probably doesn't exist in 
nature. On the other hand, the standard model has obtained tremendous 
achievement in describing electroweak interactions. So, if Higgs particle does 
not exist in nature, how to construct an electroweak model, in which Higgs 
mechanism is not used and the underlying 
dynamical  properties of the new theory are 
quite similar to those of the standard model, is an important and urgent task 
for theoretical physicists. Although we don't know whether Higgs particle 
exists in nature or not, constructing such kind electroweak model is still 
important theoretically. At least, this model gives us an important result that, 
without Higgs particle, a correct electroweak theory which coincides with 
experimental results could also be constructed. In other words, if we could 
construct such kind electroweak model, it means that Higgs particle is not a 
necessary part of an acceptable electroweak model. If in the future, physicists 
have proved that Higgs particle doesn't exist in nature, any attempt to 
construct an electroweak model without Higgs particle will become more and 
more important, for the future correct electroweak theory must come from 
this 
attempt. Therefore, from whatever point of view, constructing an electroweak 
model without Higgs particle is interesting and important theoretically. 
\\

~~~~ Although there exists no Higgs particle in the new electroweak model,
there exist a vacuum potential and some new particles in the new electroweak 
model. All these new particles are massless gauge bosons which do not exist 
in the standard model. In order to understand the roles of these massless 
gauge bosons in the new electroweak model, we first 
discuss the roles of Higgs fields in the standard model. In 
order to introduce the masses of intermediate gauge bosons, leptons and 
quarks, we must introduce Higgs scalar fields and Higgs 
mechanism in the standard model. In the 
standard model, Higgs fields have the following two important roles: 1) to 
introduce the masses of intermediate bosons, the masses of quarks and the 
masses of leptons; 2) to keep the local gauge symmetry of the original 
lagrangian so as to make the theory renormalizable. If we try to construct a 
new electroweak model in which Higgs mechanism is not used, we must look 
for some new fields which could take 
place the roles of Higgs fields in the standard 
model. Those new  massless gauge fields and vacuum potential are these 
things we are looking for. (The goal of the introduction of vacuum potential is 
to introduce symmetry  breaking and the masses of leptons, quarks and 
intermediate gauge bosons. ) In the new electroweak model, we have 
introduced two sets of gauge fields. One set of gauge fields are the original 
ones introduced in the standard  model. Anther set of gauge fields could be 
regarded as complementary fields of the original gauge fields whose function 
is to introduce the masses of the intermediate gauge bosons without breaking 
the local gauge symmetry of the original lagrangian. After two sets of fields 
transformations and symmetry breaking, one set of gauge fields obtain masses 
and another set of gauge fields keep massless. If we introduce only one set of 
gauge fields as we do in Yang-Mills theory, we could not keep the mass term 
of gauge fields local gauge invariant. We must clearly see that, in 
constructing electroweak model, it is extremely important to keep local gauge 
symmetry of the original lagrangian, for the local gauge symmetry of the 
original lagrangian will make the propagators of the massive gauge bosons
have the correct forms and 
give a Ward-Takahashi identity which will play a key 
role in the renormalization of the theory. In a word, if we want to introduce 
the masses of intermediate gauge bosons without using Higgs mechanism, 
the 
introduction of these massless gauge bosons can not be avoided, otherwise, 
the theory is non-renormalizable. In other words, in order to make the theory
renormalizable, we must keep the local gauge symmetry of the original 
lagrangian. In order to introduce the masses of intermediate gauge bosons 
without violating the local gauge symmetry of the original lagrangian, we 
must either use Higgs mechanism or introduce two sets of gauge fields in 
theory. In this paper, we avoid using Higgs mechanism, so we introduce two 
sets of gauge fields. \\

~~~~ The problem of the renormalization of the theory is mentioned several 
times above, now we will give a more detailed discussion on it. Though a 
complete strict proof on the renormalizability of the theory , which is very 
complicated and needs a relatively long time to accomplish, is not obtained 
yet, we will give some preliminary considerations on this problem. As we 
have mention before, the original lagrangian of the model has strict local gage 
symmetry. When we quantize the theory in path integral formulation, we 
should take gauge conditions first. If we take proper gauge conditions, we 
could make the propagators of massive gauge bosons have 
the following form \lbrack 11 \rbrack :
$$
\frac{i}{k^2-m^2} \left \lbrack 
-g_{\mu \nu} + ( 1- \frac{1}{\xi}) \frac{k_{\mu} k_{\nu}}{k^2 - m^2/ \xi}
\right \rbrack .
\eqno{(6.1)}
$$
In this case, it is easy to see that, according to the law of power counting, the 
general gauge field theory is a kind of renormalizable theory. So is the new 
electroweak theory. Besides, the local gauge symmetry of the original 
lagrangian will give a Ward-Takahashi identity which will eventually make 
the theory renormalizable. According to our knowledge on the 
renormalization of the standard model, we believe that the new electroweak 
theory is renormalizable. \\

~~~~ Although, in the new electroweak model, there exist massless gauge
bosons which have not been found by experiment up to now, there exists no
contradictions between high energy experiments and the new electroweak
model, because if the parameter $\alpha$ is small enough, the cross section
caused by the interchange of massless gauge bosons will be extremely small.
As an example, let's simple discuss $e^-$ - neutrino scattering. The cross
section of $e^-$ - neutrino scattering cause by the interchange of massless
gauge bosons is similar to that of Bhabha scattering whose cross section is
proportional to the fourth power of the coupling constant. According to
eq.(3.12), the coupling constant of leptons and massless gauge bosons $Z_2$ 
is:
$$
\sqrt{g^2 + {g \prime}^2} {\rm sin} \alpha .
\eqno{(6.2)}
$$
So the cross section $\sigma_1$ of $e^-$ - neutrino scattering cause by the 
interchange of massless gauge bosons is proportional to 
$$            
(g^2 + {g \prime}^2)^2 {\rm sin}^4 \alpha .
\eqno{(6.3)}
$$
The cross section  $\sigma_2$ of $e^-$ - neutrino scattering cause by the 
interchange of massive gauge bosons is proportional to 
$$                                          
\frac{g^4}{m_W^4}  {\rm cos}^4 \alpha .
\eqno{(6.4)}
$$
If
$$                                          
\alpha \sim 10^{-3} ,
\eqno{(6.5)}
$$
then 
$$
\sigma_1 \ll \sigma_2.
\eqno{(6.6)}
$$
That means that the contribution of massless gauge bosons to cross section
is much smaller than that of massive gauge bosons. Because massless gauge
bosons interact with electrons, they have contributions to the Bhabha
scattering. If $\alpha$ is in the order of $10^{-3}$, then the cross section
of $e^+ e^-$ scattering caused by massless gauge bosons is about $10^{10}$
times smaller than the cross section of $e^+ e^-$ scattering caused by
photon. Therefore, if parameter $\alpha$ is small enough, the introduction
of massless gauge bosons in the new electroweak model will cause no
inconsistency between theory and experiment.   \\

~~~~ Up to now, experimental physicists do not find massless intermediate
gauge bosons in the high energy experiments. In the new electroweak model,
there are three kinds of new particles 
which do not exist in the standard model. They
are $Z_2$ and $W_2^{\pm}$ particles. $Z_2$ is an electric neutral massless
vector particle whose properties are quite similar to those of $\gamma$
photon. Especially, the interaction properties of $Z_2$ particle are also
quite similar to those of $\gamma$ photon. The differences between $Z_2$
particle and $\gamma$ photon are: 1) the coupling constants between those
particles and matter fields are different; and 2) $Z_2$ 
particle directly interacts with neutrinos but $\gamma$ photon 
does not directly interact with neutrinos. So, we
could imagine that it will be very difficult to differentiate $Z_2$ particle
from $\gamma$ photon and to prove the existence of $Z_2$ particle directly. 
In other words, if $Z_2$ particle exists in nature,  
there must be electric neutral massless vector particles, i.e. $Z_2$
particles, mixed in $\gamma$ photons. The fact that $Z_2$ particle mixed in
$\gamma$ photon might give us a false impression that $\gamma$ photon
interacts with neutrinos directly. $W_2^{\pm}$ particles are electric
charged massless vector particles. Of cause, it is easy to differentiate
$W_2^{\pm}$ particles from $\gamma$ photon. But, because the mass of
electron is also very small, it is difficult to distinguish between
$W_2^{\pm}$ particles and electron in the mass spectrum in the high energy 
experiment. Maybe we could say
that the spins of $W_2^{\pm}$ particles and electron are different, we
could differentiate them by the information of spin. But if $W_2^{\pm}$
particles exist in nature and are produced in the experiment, we may regard
them as electron or positron suppose that a corresponding neutrino, which is
not detected by the experiment, is produced simultaneously in the
experiment, for the total spin of a system which consists of two spin
$\frac{1}{2}$ particles could be 1. So, in the high energy experiment, it
will be difficult to differentiate $W_2^{\pm}$ particles from electron
or positron. On the other hand, if we find electric charged massless vector 
particles exist in nature, we don't know what it is, because there are no
such particles in the standard model. We may think that they are  special
kind of photons, such as charged photons, or they are electrons or positrons
suppose that there are corresponding neutrinos produced in the experiment
and the errors of the measurement of masses exist. Besides, the cross
section of the production of these massless intermediate gauge bosons is
relatively very small, few massless intermediate gauge bosons are
produced in the high energy experiment. It is a
meaningful work to directly prove the existence of these massless 
intermediate gauge bosons in the experiment, for if we have proved the
existence of these massless intermediate gauge bosons in nature, it would
means that Higgs particle maybe doesn't exist in nature and it would tell us
that which electroweak model is the correct model in describing electroweak
interactions. \\

~~~~ Although we do not use Higgs mechanism in the new electroweak 
model, any one who is familiar with the standard model may have found that 
the vacuum potential $v$ is very like Higgs field. Indeed, except for the
kinematical energy terms of Higgs particle, those terms concern of vacuum 
potential in the lagrangian are completely 
the same as those of Higgs scalar fields. 
But they have essential differences. The most important difference is that, in 
the lagrangian, Higgs fields have kinematical energy terms but vacuum 
potential do not have kinematical energy terms. So there should exist a kind
of particle corresponding to the Higgs field but there exist no particle 
corresponding to the vacuum potential. This difference may give us a wrong 
impression that vacuum potential is a very heavy Higgs particle field. For, in 
the standard model, if we suppose that the mass of scalar fields is infinity, 
then the dynamical degree of freedom of Higgs field can never be excited. So 
we could let:
$$
\partial_{\mu} \Phi \simeq 0.
\eqno{(6.7)}
$$
This opinion is not correct, because, in the standard model, the coefficient
of the mass term of the scalar field $\Phi$ in the 
origianl lagrangian is negative, we could not think that 
the mass of the scalar field is infinity and could not let all $\partial_{\mu} 
\Phi$ vanish. Therefore, vacuum potential can not be regarded as a very 
heavy Higgs field. Besides, in the standard model, 
if we let $\partial{\mu} \Phi$ vanish, then 
the local gauge symmetry is broken and the theory will be non-
renormalizable. 
Similarly, if we add kinematical energy  terms of vacuum potential to the
lagrangian of the new electroweak model, then the 
lagrangian  will lose local gauge symmetry and 
the theory will be non-renormalizable too. Therefore, though vacuum 
potential and Higgs particle have similar characteristics, they have essential 
differences.  \\

~~~~ What is vacuum? In quantum field theory, vacuum is regarded as the 
ground state which has the lowest energy of the system. In a point of view, 
vacuum is regarded as a spin-0 scalar field whose 4-momentum is always zero 
in any condition. It is a special kind of media. In the new electroweak theory, 
vacuum is regarded as a scalar field which has no dynamical degree of 
freedom. Because vacuum potential has no dynamical degree of freedom, it 
carries no energy-momentum. But it could carry some quantum numbers. It 
serves as a background in which all matters in universe  move and evolve, 
but vacuum itself can not be excited or move. All fields will interact with 
vacuum when they exist and evolve in vacuum, which had already been expressed in 
the lagrangian of the new electroweak model. For a quantum system, not only 
the properties of vacuum affect the dynamical behavior of the system, but also 
the symmetry of vacuum determines the symmetry of the system. So, when 
the 
symmetry of vacuum breaks, the symmetry of our physical world is broken 
simultaneously. In this paper, we use $v$ to represent the influence of the 
vacuum, and the symmetry breaking of the system is caused by the symmetry 
breaking of vacuum potential $v$. After symmetry breaking, $v$ has definite 
value,  which means that, in our universe, the vacuum is uniform and the 
properties of vacuum is stable. In the standard model, the masses of all fields,
include quark fields, lepton fields and gauge fields, are generated from
their interactions with Higgs field. Now, we could think that the masses of
all fields are generated from their interactions with vacuum.  This view of 
point coincides with the interaction picture of the perturbation theory. For 
example, in the perturbation theory, the self-energy diagram of electron will 
change the mass of electron and the self-energy diagram is generated from the 
interactions between vacuum and electron, for  if there were no vacuum, 
there would be no self-energy diagram. So it is natural to think that the 
masses 
of all fields are generated from their interactions with vacuum.  \\

~~~~ Because there exists electric-neutral massless intermediate gauge boson
$Z_2$ which could transmit a long-range force field, there will exist a new
long-range force in the new electroweak model. Because the coupling 
constant of massless intermediate boson field and matter fields is very small, 
the corresponding long-range force field will be very weak and its 
macroscopic effects are hard to be detected. Because neutrinos carry weak 
charge and macroscopic objects could absorb neutrinos, any macroscopic 
object will tend to be in a state of weak charge neutral.  This will make the 
macroscopic effects of weak long-range force  weaker and make the effects 
of weak long-range force field harder to be detected. If $\alpha$ is about 
$10^{-3}$, then the weak long-range force is about one million times 
weaker 
than electromagnetic force. Generally speaking, except for neutrinos, an 
object which carries weak charge will also carries electric charge, and 
electromagnetic interactions are much stronger than weak long-range 
interactions, so , we could imagine that it would be extremely difficult to find 
macroscopic effects of weak long-range force. Because electron and proton 
carry not only electric charge but also weak charge, there are weak long-range 
interactions mixed in the traditional electromagnetic interactions and the 
weak 
long-range interactions will contribute a extremely small part to the spectrum 
of atoms. Weak long-range force may have some influences on cosmology, 
for  neutrinos carry weak  charge and a huge amount of neutrinos exist in 
universe. \\

~~~~ We know that, using Higgs mechanism, we could construct a lot of
standard model. But we could construct only one electroweak model by using
vacuum potential. The reason is that there is only one vacuum potential
corresponding to a symmetry, but we could introduce many Higgs fields
corresponding to a symmetry. So, in the new theory, more parts of the 
lagrangian of the model are fixed by the symmetry. This characteristic is
important in theory. \\

~~~~ Although we could construct an electroweak model without using 
Higgs
mechanism, this does not mean that Higgs particle does not exist in nature.
In this paper, we only want to point out 
one important thing that, without Higgs particle,
we could also construct a correct electroweak model in theory. Whether 
Higgs
particle exists in nature or not should be determined by experiment. But if
we find that massless gauge bosons exist in nature, we will say that Higgs
particle maybe does not exist in nature. \\

\section*{Reference:}
\begin{description}
\item[\lbrack 1 \rbrack]  C.N.Yang, R.L.Mills, Phys Rev {\bf 96} (1954) 
191
\item[\lbrack 2 \rbrack]  T.D.Lee, M. Rosenbluth, C.N.Yang, Phys Rev {\bf
75}(1949): 9905
\item[\lbrack 3 \rbrack]  S.Glashow, Nucl Phys {\bf 22}(1961) 579
\item[\lbrack 4 \rbrack]  S.Weinberg, Phys Rev Lett {\bf 19} (1967) 1264
\item[\lbrack 5 \rbrack]  A.Salam, in Elementary Particle Theory, 
eds.N.Svartholm(Almquist and Forlag, Stockholm,1968)
\item[\lbrack 6 \rbrack]  Particle Data Group, phys. Rev. {\bf D54} (1996) 
20
\item[\lbrack 7 \rbrack]  Ning Wu, Gauge Field Theory With Massive 
Gauge Bosons, hep-ph/9802236
\item[\lbrack 8 \rbrack]  Ning Wu, General Gauge Field theory,
hep-ph/9805453
\item[\lbrack 9 \rbrack]  Ning Wu, A new electroweak model, hep-
ph/9802237
\item[\lbrack 10 \rbrack]  M.Kabayashi and M.Maskawa, Prog. Theor. 
Phys. {\bf 49}  (1973): 652
\item[\lbrack 11 \rbrack]  Ning Wu, The quantization of the general gauge 
field theory, (in preparation)
\end{description}

\end{document}